\documentclass[aps,prb,floats,preprintnumbers,superscriptaddress]{revtex4}
\usepackage{amsmath}
\usepackage{amsfonts}
\usepackage{amssymb}
\usepackage{color}
\usepackage{feynmf}
\usepackage{bm}
\usepackage{ dsfont }
\newcommand{\beq}{\begin{equation}}
\newcommand{\eeq}{\end{equation}}
\newcommand{\beqa}{\begin{eqnarray}}
\newcommand{\eeqa}{\end{eqnarray}}
\usepackage{graphicx,amssymb,amsmath,subfigure}
\usepackage{bm}

\begin{document}
\bibliographystyle{naturemag}

\title{Strain induced Chiral Magnetic Effect in Weyl semimetals}

\author{Alberto Cortijo$^{1}$, Dmitri Kharzeev$^{2,3}$, Karl Landsteiner$^{4}$ and Maria A.H. Vozmediano$^{}$}
\email{$^{1}$Alberto.Cortijo@csic.es, $^{2}$Dmitri.Kharzeev@stonybrook.edu, $^{4}$Karl.Landsteiner@csic.es, $^{5}$vozmediano@icmm.csic.es}
\affiliation{
$^{}$ Instituto de Ciencia de Materiales de Madrid, C/ Sor Juana In\'es de la Cruz 3, 
Cantoblanco, 28049 Madrid, Spain\\
${^2}$Department of Physics and Astronomy, Stony Brook University, Stony Brook, New York 11794-3800, USA\\
$^{3}$Department of Physics and RIKEN-BNL Research Center, Brookhaven National Laboratory, Upton, New York 11973-5000, USA\\
$^{4}$Instituto de F\'{\i}sica Te\'orica UAM/CSIC, C/ Nicol\'as Cabrera 13-15, Cantoblanco, 28049 Madrid, Spain
}

\begin{abstract}
We argue that strain applied to a time-reversal and inversion breaking Weyl semi-metal in a magnetic field can induce an electric current via the chiral magnetic effect.
A tight binding model is used to show that strain generically changes the locations in the Brillouin zone but also the energies of the band
touching points (tips of the Weyl cones). Since axial charge in a Weyl semi--metal can relax via inter-valley scattering processes the induced current will decay with a timescale given by the
lifetime of a chiral quasiparticle. We estimate the strength and lifetime of the current for typical material parameters and find that it should be experimentally observable.
\end{abstract}

\preprint{IFT-UAM/CSIC-16-065}
\maketitle
%
Chiral anomalies belong to the most emblematic predictions of quantum field theory
\cite{Bert96}. In the presence of massless (chiral) fermions not all
the symmetries
of the classical theory are compatible with the quantum theory. The theory of
anomalies has a very wide range of applications reaching
from particle physics to condensed matter physics. 
In the recent years anomaly
induced transport phenomena, such as the chiral magnetic effect (CME)
have attracted much attention in the quark--gluon plasma
\cite{KZ07,KMW08,FKW08}, and in Dirac matter \cite{BGKY14,LKetal16}. The recently synthesized Weyl semimetals (WSM) \cite{XuLiuetal15,Xu2etal15,LVetal15} are an optimum benchmark to test anomaly related phenomena in condensed matter. These materials have chiral Weyl points separated in momentum space. The minimal model with two Weyl nodes in a continuum description is described by the action
\begin{equation}
\label{eq:Weyl}
S=\int d^4k \bar{\psi}_{-k}(\gamma^{\mu}k_{\mu}-m-b_{\mu}\gamma^{\mu}\gamma_{5})\psi_{k}.
\end{equation}
which resembles a Lorentz breaking QED action \cite{Gr12,VF13}. The WSM phase is reached when the parameters obey the condition $-b^2 > m^2$ in which case the separation between nodes  is proportional to the four-vector 
$\Delta k^{\mu} \equiv \lambda^\mu\sim b^{\mu}\sqrt{1-\frac{m^2}{b^2}}$.

The CME
describes the generation of an electric current parallel to an applied magnetic field
in chirally imbalanced matter \cite{KZ07,KMW08, FKW08}. It the context of Weyl semi-metals it is best described by the formula \cite{Gynther:2010ed,BGKY14}
\begin{equation}\label{eq:cme}
 \vec{J}  = \frac{\mu_L -\mu_R + E_L - E_R}{4\pi^2} \vec{B} = \frac{\mu_5 - \lambda_0}{2\pi^2} \vec B\,,
\end{equation}
where $\mu_{L,R}$ are the chemical potentials of  left- and right-handed fermions as measured from the tips of the Weyl cones whereas $E_{L,R}$ are the 
energies at which the Weyl cones are located. The axial chemical potential is $2 \mu_5 = \mu_L-\mu_R$ and the difference in energy of the  band touching
points is $2\lambda_0 = E_R-E_L$.

The axial symmetry is not exact. It is broken even at tree level by the mass term in  (\ref{eq:Weyl}). In a crystal this is unavoidable due to the compactness of the
Brillouin zones. Therefore axial charge will decay to a thermodynamic equilibrium state in which the left- and right-handed fermi surfaces sit at the same
energy such that $E_L+\mu_L = E_R+\mu_R$. The CME vanishes in a Weyl semi-metal in thermodynamic equilibrium \cite{VF13,BGKY14}. 

In order to induce a non-vanishing chiral magnetic current it is clear from (\ref{eq:cme}) that there are in principle two options. First one might modify the chemical potentials by 
inducing an imbalance in the occupation numbers of left- and right-handed fermions. One way to do this is to use the proper axial anomaly
\begin{equation}
 \partial_\mu J^\mu_5 = \frac{1}{2\pi^2} \vec{E}\cdot\vec{B}\,.
\end{equation}
In parallel electric and magnetic fields, fermions are pumped via spectral flow from one Weyl-cone to the other and thereby changing their chirality. 
This process leads to a dramatic enhancement of the electric conductivity along the direction of the magnetic field. The anomaly induced negative magnetoresitivity
has indeed been verified experimentally in Dirac and Weyl semi-metals \cite{ZXetal16,LHetal16,Shetal15b}.

A second possibility is to change the locations of the Weyl cones $E_{L,R}$. 
We argue in the reminder of this work 
that this can indeed be achieved by applying strain, in particular, it will be induced by the time component of the axial elastic vector field that emerges when lattice deformations are included in the model. As a result, we find an elastic contribution to the chiral magnetic effect linearly proportional to the strain tensor $u_{ij}$ and to the Gr{\"u}neisen parameter of the material $\beta$.

\noindent
{\it Strain induced chiral imbalance}

\begin{figure}
\includegraphics[width=18cm]{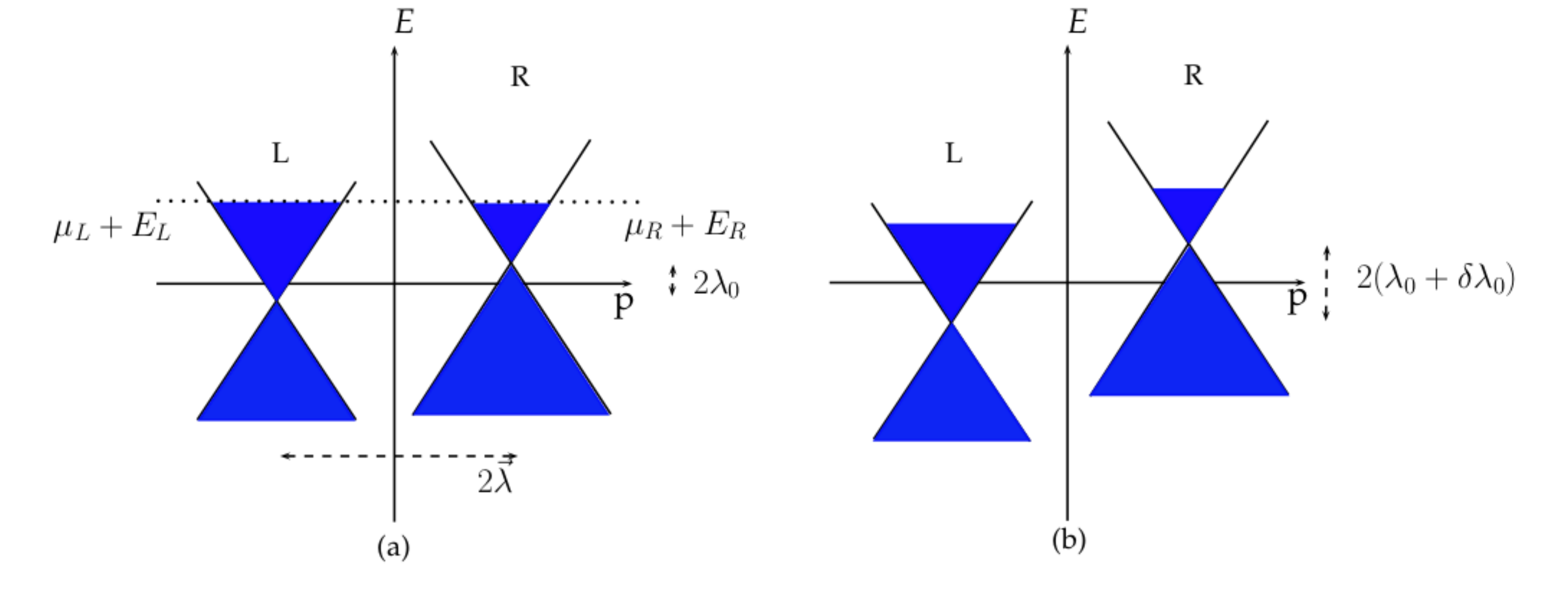}
\caption{Local band structure of a Weyl semimetal with two nodal points at different energies and different position in momentum space. 
In (a) there are Fermi surfaces of different sizes for left- and right-handed fermions as measured by the left- and right-handed chemical potentials $\mu_{L,R}$.
The Fermi energy of both Weyl cones is however the same in equilibrium resulting in vanishing CME. In (b) the situation shortly after applying strain is
depicted. The tips of the Weyl cones have shifted in energy but the Fermi surfaces have not yet had enough time time to equilibrate $t<\tau_\mathrm{inter}$.
The CME is given by $\vec{J} = \frac{\delta\mu_5(t) - \delta b_0}{2\pi^2} \vec B$.  Over a timescale given by the inter valley scattering time $\tau_\mathrm{inter}$
the axial chemical potential $\delta \mu_5$ will build up such at for for $t\gg \tau_\mathrm{inter}$ it takes the value $\delta \mu_5 = \delta b_0$ and equilibrium with vanishing 
CME is reached again.}
\label{fig:Weyl}
\end{figure}  
The local band structure  for a WSM with two Weyl points is described by the continuum model
\begin{equation}
\label{eq:Weyl2}
S=\int d^4k \bar{\psi}_{k}(\gamma^{\mu}k_{\mu}-\lambda_{\mu}\gamma^{\mu}\gamma_{5})\psi_{k}.
\end{equation}
The schematic band structure is represented in Fig. \ref{fig:Weyl}. The space components of the  vector 
$\lambda_\mu$ breaks time reversal symmetry and its magnitude sets  the separation of the Weyl nodes in momentum space. Its time component $\lambda_0$ breaks inversion symmetry and marks the separation in energy of the points.
The emergence of elastic gauge fields coupling with opposite signs to the two chiralities (axial) in time reversal breaking Weyl semimetals was derived from a tight binding model in ref. \onlinecite{CFLV15} (see also \onlinecite{CZ16}). In Appendix \ref{sec_TB} we  extend this derivation to account for a time component of the elastic gauge field. 
The elastic gauge field shifts the locations of the Weyl tips and induce an effective axial gauge field 
\begin{equation}
 2(\delta \lambda_0) =  (E_L-E_R) \beta u_{33}\,.
\end{equation}

%
The strain induced CME is proportional to the difference in energies of the Weyl-cones before applying strain, to the Gr{\"u}neisen parameter of the material, and to the magnitude of the
strain itself. Since eventually the Fermi surfaces will equilibrate by processes such as inter--valley scattering or through spectral flow through the edge states (Fermi arcs),
this strain induced CME will be observable only over a short timespan set by the lifetime $\tau_\mathrm{inter}$ of a chiral quasiparticle at the corresponding Fermi surface. This time is estimated to be much larger than the intra-valley scattering rate or the usual impurity induced lifetime \cite{XKetal15,BGetal16}.
Taking anomaly and inter--valley scattering into account the time developement of the axial chemical potential is given by
\begin{equation}
 \partial_t \rho_5 = \frac{1}{2\pi^2} \vec{E}.\vec{B} - \frac{1}{\tau_\mathrm{inter}} \left(\rho_5 - \rho_5^{(eq)}\right)
\end{equation}
where $\rho_5^{(eq)}= \rho_5(\lambda_0+\delta\lambda_0)$.
The relation between the axial chemical potential and the axial charge is given by the equation of state $\rho_5 = \frac{\partial P}{\partial \mu_5}$ with $P$ the
the thermodynamic pressure for chiral fermions. This will result in an exponentially decaying strain induced chiral magnetic current
\begin{equation}
 \vec{J} = \frac{\delta \lambda_0}{2 \pi^2} \vec B e^{- t/\tau_{\mathrm{inter}}} = \frac{(E_L-E_R)\beta u_{33}}{4 \pi^2} \vec B e^{- t/\tau_{\mathrm{inter}}}\,.
\end{equation}

\noindent
{\it Physical feasibility of the proposal}
\\

We can estimate the strength of the current by taking the typical separation of the Weyl nodes in energy be of the order of $1$ meV, the Gr{\"u}neisen parameter $\beta=1$ and
the attainable amount of strain in the percentage realm ($1$\% ). The resulting current is \begin{equation}
 {\bf J} \approx 10^{5}{\bf B} \; A/m^2, 
\end{equation}
with the magnetic field  $B$ expressed in Tesla. For a sample of a cross section of $10 \mu m^2$ the induced current is of the order of 10 microamperes in a magnetic field of 1T.
This an observable quantity of the same order of magnitude as the one described in \cite{SN16,BHTB16}. Since the considered WSM materials break inversion symmetry, they will also be piezoelectric.  For comparison, the maximum current in a typical piezoelectric  varies from $nA$ to $\mu A$ and the voltage generated in $1-100 V$, depending on the size of the material. The CME signal should be easily distinguishable from the standard piezo response due to its linear dependence on the magnetic field.
The current will decay exponentially within a typical time  $10 \tau_{inter}$. 

For the expression (\ref{eq:cme}) to be applicable we need to assume that the strain induced deformation acts on a time scale which is slow compared to the equilibration
time with a Weyl cones but faster than the equilibration time for the axial charge, i.e. the inter valley scattering time. This can be fulfilled in WSM where
inter valley relaxation time is estimated to be at least two orders of magnitude larger than  standard lifetime \cite{XKetal15}.

A potential difficulty to measure the effect lies on the fact that the inter-valley relaxation time (estimated to be of the order of $10^{-9}$ s at best \cite{BGetal16}) is of the same order of magnitude than the time needed for the elastic perturbation to propagate along the sample (assumed of microns size) what might attenuate the signal. Still the magnitude of the induced current will be within the observable range.

The class of materials realizing the WSM physics is growing exponentially in time. The expectations are that we will be able to get WSM {\it \`a la carte} according to the theoretical construction described in \cite{CSetal16,CBetal16}.
Effects of strain in WSM has been considered previously in various contexts \cite{BTR15,RJetal16} but the contribution of elastic gauge fields is novel and potentially more relevant.

\acknowledgments
MAHV thanks F. de Juan and A. Grushin for useful conversations.
The work of A. C. and M.A.H.V.  has been supported by Spanish MECD
grants FIS2014-57432-P, the European
Union structural funds and the Comunidad de Madrid
MAD2D-CM Program (S2013/MIT-3007), and by the European Union Seventh Framework
Programme under grant agreement no. 604391 Graphene Flagship.
The work of D. K. has been supported by in part by the U.S. Department of Energy under Contracts No. DE-FG-
88ER40388 and DE-AC02-98CH10886.
The work of K.L. has bee supported by Severo Ochoa Programme grant SEV-2012-0249 and by FPA2015-65480-P (MINECO).
K.L. and M.A.H.V. gratefully acknowledge support from the Simons Center for Geometry and Physics, Stony Brook University at which  some of the research for this paper was performed.
\bibliography{CME}
\appendix
\section{Time component of the elastic gauge field}
\label{sec_TB}
\subsection{The model with $b_0$}
The minimal tight binding model reproducing the band structure of WSM is that of of s-, and p-like electrons hopping in a cubic lattice and chirally coupled to an on-site constant vector field $b_\mu$\cite{VF13,SHR15}:
\beqa
H_{0}&=&\sum_{i}\sum_{s}\frac{-1}{2}c^{+}_{i}\left(r\hat{\beta}+i t \alpha_{s}\right)c_{i+s}-\frac{1}{2}c^{+}_{i+s}\left(r\hat{\beta}-i t \alpha_{s}\right)c_{i}+\sum_{i}\Delta\; c^{+}_{i}\hat{\beta} c_{i}\\ \nonumber
&+&\sum_{i}b_{3}c^{+}_{i}\alpha_{3}\gamma_{5}c_{i}
+\sum_{i}b_{0}c^{+}_{i}\alpha_{3}c_{i},\label{Ham1}
\eeqa
where $t, r, m$ are hopping parameters between $s$ and $p$ states, hopping between the same kind of states, and the difference of on-site energies between $s$ and $p$ states, respectively. The parameter $\Delta$ is $\Delta=m+3r$. Without loss of generality, we choose the vector field $\bm{b}$ to point along the OZ direction $b_3$.
This spacial component breaks time reversal  $\mathcal{T}$ and $SO(2)$ rotational symmetry; the time component $b_0$ shifts the Weyl cones in energy and  breaks inversion symmetry.  
In momentum space,
\beqa
H_{0}=\sum_{\bm{k},s}t \sin(k_{s}a)c^{+}_{\bm{k}}\alpha_{s}c_{\bm{k}}-r\sum_{\bm{k},s}\cos(k_{s}a)c^{+}_{\bm{k}}\hat{\beta} c_{\bm{k}}+\Delta\sum_{\bm{k}}c^{+}_{\bm{k}}\hat{\beta} c_{\bm{k}}+b_{3}\sum_{\bm{k}}c^{+}_{\bm{k}}\alpha_{3}\gamma_{5}c_{\bm{k}}+b_{0}\sum_{\bm{k}}c^{+}_{\bm{k}}\gamma_{5}c_{\bm{k}}.\label{Hamk}
\eeqa
with $s=1,2,3$. We will use the following set of Dirac matrices, $\alpha_{1}=\tau_{0}\sigma_{1}$, $\alpha_{2}=\tau_{0}\sigma_{2}$, $\alpha_{3}=\tau_{1}\sigma_{3}$, $\hat{\beta}=\tau_{3}\sigma_{3}$, $\gamma_{5}=\tau_{1}\sigma_{0}$, so $\alpha_{3}\gamma_{5}=\tau_{0}\sigma_{3}$.

We will choose to expand the Hamiltonian (\ref{Hamk}) around the $\bm{k}=0$ point ($\bm{k}\cdot\bm{p}$ theory) and compute the value of the emergent Fermi points within this approximation. When computing the effect of the strain we will start from this approximation. Around the $\Gamma$ point, $\sin(k_{s}a)\simeq k_{s}a$, and $\cos(k_{s}a)\simeq 1$, so the Hamiltonian matrix reads ($v=ta$, and $m=\Delta-3r$)

\beqa
\mathcal{H}_{0}(\bm{k})=\left(\begin{array}{cc}
v\bm{\sigma}\cdot\bm{k}_{\perp}+(m+b_{3})\sigma_{3} & vk_{3}\sigma_{3}+b_0 \sigma_0\\
vk_{3}\sigma_{3}+b_0\sigma_0 & v\bm{\sigma}\cdot\bm{k}_{\perp}+(b_{3}-m)\sigma_{3}
\end{array}\right),
\eeqa
acting on the spinor $\Psi_{\bm{k}}=(\phi_{\bm{k}},\psi_{\bm{k}})^{T}$. The momentum $\bm{k}_{\perp}$ is the momentum perpendicular to $b_{3}$. The main difference with respect to the original case is that now, for energies $\omega, v|\bm{k}_{\perp}|\ll m+b_{3}$, the high energy sector represented by $\phi_{\bm{k}}$ is related to $\psi_{\bm{k}}$ through
\beq
\phi_{\bm{k}}\simeq-\frac{\sigma_0 v k_{3}+\sigma_3 b_0}{m+b_{3}}\psi_{\bm{k}},\label{highsector}
\eeq
 and the effective two-band model is

\beq
H_{eff}=\sum_{\bm{k}}\psi^{+}_{\bm{k}}\left(-\frac{2b_0 vk_3}{m+b_3}\sigma_0+v\bm{\sigma}\cdot\bm{k}_{\perp}+\frac{1}{m+b_{3}}(b^{2}_{3}-b^2_0-m^{2}-v^{2}k^{2}_{3})\sigma_{3}\right)\psi_{\bm{k}}.
\eeq

The presence of $b_0$ induces two differences: first, there is a term proportional to the identity matrix $\sigma_0$ and \emph{linearly} dependent on $k_3$ breaking inversion symmetry (in the original full lattice model, it would be proportional to $\sin(k_3 a)$), and second, it modifies the mass term accompanying the $\sigma_3$ matrix. 

The condition to obtain the Weyl nodes is now when the two bands intersect each other: $E_+(\bm{k})=E_{-}(\bm{k})$, so the positions of the Weyl nodes change correspondingly to:

\beq
\lambda_{\pm}=\pm \lambda_3=\pm\frac{\sqrt{b^{2}_{3}-b^2_0-m^{2}}}{v},
\eeq
that is, when $b_0=0$, the two Weyl nodes appeared for values of $b_3>m$. Now, the situation is a little bit more complex (and richer), since for some values of the three parameters $m, b_3$, and $b_0$ it might happen that the maximum of the valence band is placed at higher energies than the minimum of the conduction band, but the two bands do not intersect at any real momentum. The energies where the new Weyl nodes are placed read:
\beq
E_{\pm}=\mp2b_0\frac{\sqrt{b^2_3-b^2_0-m^2}}{m+b_3}.
\eeq
Expanding around $\lambda_3$, $k_{3}=\pm\lambda_3+\delta k_{3}$, for small $\delta k_{3}$ the Hamiltonian matrix around the two Fermi points take the form ($\tau=\pm1$ labels the two Weyl points)

\beq
H^{W}_{\tau}(\bm{k})=\tau \frac{2vb_0\lambda_3}{m+b_3}\sigma_0+v\bm{\sigma}\cdot\bm{k}_{\perp}+\tau v_{3}(k_{3}+\tau\lambda_3)\sigma_{3},
\eeq
that is, the Hamiltonian of two Weyl modes with opposite chirality $\big(v_{3}=2v\sqrt{\frac{b^2_{3}-b^2_0-m^2}{(b_{3}+m)^2}}\big)$.
\subsection{Elastic deformations}
Elastic deformations of the lattice induce
 the following two changes in the Hamiltonian (\ref{Hamk}):

\begin{subequations}
\beq
t\alpha_{s}\rightarrow t(1-\beta u_{ss})\alpha_{s}+t\beta\sum_{s'\neq s}u_{ss'}\alpha_{s'},
\eeq
\beq
r\rightarrow r_{s}=r(1-\beta u_{ss}).
\eeq
\label{recipe}
\end{subequations}
The tensor $u_{ij}$ is the strain tensor. We will make the approximation of setting all the Gruneisen parameters $\beta$ to be equal. After expanding first around the Weyl points and applying these changes in (\ref{Hamk}) we can split the new Hamiltonian into $H_{0}+H[u]$:

\beq
H[u]=-\beta v\tau\lambda_3\sum_{k}u_{33}  c^{+}_{\bm{k}}\alpha_{3}c_{\bm{k}}+\beta v \tau\lambda_3\sum_{s\neq 3}u_{3s} c^{+}_{\bm{k}}\alpha_{s}c_{\bm{k}}+r\beta\sum_{\bm{k}}Tr[u]c^{+}_{\bm{k}}\hat{\beta}c_{\bm{k}}.\label{Hamu}
\eeq

Remembering the expression (\ref{highsector}) relating the high energy sector to the low energy sector, we can evaluate the elements of the type $c^{+}_{\bm{k}}\alpha_{s}c_{\bm{k}}$ keeping the lowest order in a $1/(m+b_3)$ expansion:
\begin{subequations}
\beq
c^{+}_{\bm{k}}\alpha_{3}c_{\bm{k}}\simeq \phi^{+}_{\bm{k}}\sigma_3\psi_{\bm{k}}+\psi^{+}_{\bm{k}}\sigma_3\phi_{\bm{k}}=-\frac{2}{m+b_3}\psi^{+}_{\bm{k}}(\tau v\lambda_3\sigma_3+b_0\sigma_0)\psi_{\bm{k}},
\eeq
\beq
c^{+}_{\bm{k}}\alpha_{s}c_{\bm{k}}\simeq \psi^{+}_{\bm{k}}\sigma_s \psi_{\bm{k}},
\eeq
\beq
c^{+}_{\bm{k}}\hat{\beta}c_{\bm{k}}\simeq-\psi^{+}_{\bm{k}}\sigma_3 \psi_{\bm{k}}.
\eeq
\end{subequations}
With these expressions, we the have
\beqa
H[u]&=&\sum_{k}\left(2\tau b_0\beta\frac{v\lambda_3}{m+b_3}u_{33}\right)\psi^{+}_{\bm{k}}\sigma_0 \psi_{\bm{k}}+\sum_{k,s=1,2}\left(\tau\beta v\lambda_3 u_{3s}\right)\psi^{+}_{\bm{k}}\sigma_s \psi_{\bm{k}}+\nonumber\\
&+&\sum_{k}\left(2\beta \frac{v^2\lambda^2_3}{m+b_3}u_{33}-\beta r Tr[u]\right)\psi^{+}_{\bm{k}}\sigma_3 \psi_{\bm{k}},\label{Hvector}
\eeqa
from which we can read the components of the elastic vector field:
\begin{subequations}
\beq
A^{el}_{0}=\tau b_0\beta\frac{2v\lambda_3}{m+b_3}u_{33},
\eeq
\beq
A^{el}_{1}=\tau\beta v\lambda_3 u_{31},
\eeq
\beq
A^{el}_{2}=\tau\beta v\lambda_3 u_{32},
\eeq
\beq
A^{el}_{3}=2\beta \frac{v^2\lambda^2_3}{m+b_3}u_{33}-\beta r Tr[u].
\eeq
\end{subequations}
The novelty now is that, because the presence of the term $b_0$ in the original hamiltonian, a chiral zeroth component of the elastic vector field appears. 

The same derivation can be obtained from general symmetry arguments along the  lines of ref. \cite{MJSV13}. We note that, from a general symmetry approach,  even if the unstrained material has $b_0=0$,  an effective coupling of the form $\tau A_0(x)\sigma_0$ will be generated in the low energy effective action since a generic strain deformation will break inversion symmetry. Moreover, since rotational invariance is broken in the material by the vector ${\bf b}$, the only symmetry remaining is rotations in the perpendicular plane. If ${\bf b}$ points, say, along the OZ axis, the $u_{33}$ component of the strain tensor is a scalar quantity that can be coupled to any term in the effective Hamiltonian.

\end{document}